\documentclass[12pt,preprint]{aastex}
\usepackage{natbib}

\newcommand{\arcsecpoint}{\ifmmode ''\!. \else $''\!.$\fi}

\newcommand{\kms}{\ifmmode {\rm km\ s}^{-1} \else km s$^{-1}$\fi}
\newcommand{\Msun}{\ifmmode {\rm M}_{\odot} \else M$_{\odot}$\fi}
\newcommand{\Lsun}{\ifmmode {\rm L}_{\odot} \else L$_{\odot}$\fi}
\newcommand{\qo}{\ifmmode q_{\rm o} \else $q_{\rm o}$\fi}
\newcommand{\Ho}{\ifmmode H_{\rm o} \else $H_{\rm o}$\fi}
\newcommand{\ho}{\ifmmode h_{\rm o} \else $h_{\rm o}$\fi}
\newcommand{\ltsim}{\raisebox{-.5ex}{$\;\stackrel{<}{\sim}\;$}}
\newcommand{\gtsim}{\raisebox{-.5ex}{$\;\stackrel{>}{\sim}\;$}}
\newcommand{\vFWHM}{\ifmmode v_{\mbox{\tiny FWHM}} \else
                    $v_{\mbox{\tiny FWHM}}$\fi}
\newcommand{\CCF}{\ifmmode F_{\it CCF} \else $F_{\it CCF}$\fi}
\newcommand{\ACF}{\ifmmode F_{\it ACF} \else $F_{\it ACF}$\fi}
\newcommand{\Halpha}{\ifmmode {\rm H}\alpha \else H$\alpha$\fi}
\newcommand{\Hbeta}{\ifmmode {\rm H}\beta \else H$\beta$\fi}
\newcommand{\Hgamma}{\ifmmode {\rm H}\gamma \else H$\gamma$\fi}
\newcommand{\Hdelta}{\ifmmode {\rm H}\delta \else H$\delta$\fi}
\newcommand{\Lya}{\ifmmode {\rm Ly}\alpha \else Ly$\alpha$\fi}
\newcommand{\Lyb}{\ifmmode {\rm Ly}\beta \else Ly$\beta$\fi}
\newcommand{\HeI}{\ifmmode {\rm He}\,{\sc i}\,\lambda5876 \else 
	          He\,{\sc i}\,$\lambda5876$\fi}
\newcommand{\HeII}{\ifmmode {\rm He}\,{\sc ii}\,\lambda4686 \else 
	           He\,{\sc ii}\,$\lambda4686$\fi}

\newcommand{\fe}{Fe}
\newcommand{\feii}{Fe\,{\sc ii}}
\newcommand{\feiii}{Fe\,{\sc iii}}
\newcommand{\neiii}{Ne\,{\sc iii}}
\newcommand{\neiv}{Ne\,{\sc iv}}

\newcommand{\cii}{C\,{\sc ii}}
\newcommand{\ciii}{\ifmmode {\rm C}\,{\sc iii} \else C\,{\sc iii}\fi}
\newcommand{\civ}{\ifmmode {\rm C}\,{\sc iv} \else C\,{\sc iv}\fi}

\newcommand{\oiii}{O\,{\sc iii}}

\newcommand{\ovi}{O\,{\sc vi}}

\newcommand{\mgii}{Mg\,{\sc ii}}

\begin{document}

\slugcomment{To be published in Astrophys.Journal}

%
%

\title{High Redshift Quasars and Star Formation in the Early Universe
       {$^\ast$}\\ 
}

\author{M.\,Dietrich
\altaffilmark{1,2}
I.\,Appenzeller
\altaffilmark{2}
M.\,Vestergaard
\altaffilmark{3}
S.J.\,Wagner
\altaffilmark{2}
}
\altaffiltext{1}
{Department of Astronomy, University of Florida, 211 Bryant Space Center, 
 Gainesville, FL 32611-2055, USA}   
\altaffiltext{2}
{Landessternwarte Heidelberg--K\"onigstuhl, K\"onigstuhl 12, 
 D--69117 Heidelberg, Germany}
\altaffiltext{3}
{Department of Astronomy, The Ohio State University, 140 West 18th Av.,
 Columbus, OH 43210-1173, USA\\[1mm]
 $^\ast$ Based on observations collected at the European Southern 
         Observatory, La Silla, Chile}
\email{dietrich@astro.ufl.edu}
%
%

\begin{abstract}
In order to derive information on the star formation history in the early 
universe we observed 6 high-redshift (z$\simeq$3.4) quasars in the 
near-infrared to measure the relative iron and \mgii\ emission strengths. 
A detailed comparison of the resulting spectra with those of low-redshift
quasars show essentially the same \feii /\mgii\ emission ratios and very 
similar continuum and line spectral properties, indicating a lack of evolution
of the relative iron to magnesium abundance of the gas since 
$z\simeq 3.4$ in bright quasars.
On the basis of current chemical evolution scenarios of galaxies, where 
magnesium is produced in massive stars ending in type II SNe, while iron is 
formed predominantly in SNe of type Ia with a delay of $\sim 1$\, Gyr 
and assuming as cosmological parameters H$_o = 72$ km\,s$^{-1}$\,Mpc$^{-1}$, 
$\Omega _M = 0.3$, and $\Omega _\Lambda = 0.7$,  we 
conclude that major star formation activity in the host galaxies of our 
$z\simeq 3.4$ quasars must have started already at an epoch corresponding  
to $z_f \simeq 10$, when the age of the universe was less than $0.5$ Gyrs.
\end{abstract}

\section{Introduction}
Quasars are among the most luminous objects in the universe. Because of their 
high luminosity, bright quasars can be observed at practically any distance
at which these objects are expected to occur. Among the characteristic
properties of quasars are prominent emission lines from the gas ionized
and heated by a central radiation source. Assuming that this gas originates 
from the interstellar medium of the quasar host galaxies, line diagnostics 
applied to the prominent emission lines provide an important tool to study the 
chemical composition and enrichment history of the quasar host galaxies at
all cosmological epochs. Earlier studies of the emission line spectra have 
demonstrated that quasars at redshift $z\geq 3$ have metallicities of up 
to an order of magnitude larger than the solar value 
\citep[e.g.,][]{HaFe92,HaFe93,HaFe99,Feetal96,Koetal96,pet99,Dietal99,DiWi00}.
These results show that well before the epoch corresponding to $z = 3$ 
significant star formation must have taken place in the galactic or 
proto-galactic cores where these quasars reside.

Some information on the exact epoch of the first star formation activity in 
these objects can be derived from the relative abundance of $\alpha $-process 
elements and iron. According to present chemical enrichment scenarios 
$\alpha $-element nuclei are produced predominantly in type II SNe with 
massive progenitors on time scales of 
$\tau _{evol}\simeq 2 \cdot 10^6$ -- $10^7$ 
years after the beginning of the star formation epoch.
On the other hand, the dominant source of iron is assumed to be type Ia SNe at
the end point of the evolution of intermediate mass stars in binary systems, 
about $\tau _{evol}\simeq 1$\,Gyr after the onset of the star formation epoch
\citep[e.g.,][]{Tins79,MaGr86,WST89,Yoetal96}.
The amount of iron returned to the interstellar medium in SN\,II ejecta is 
rather low \citep{Yoetal96,Yoetal98}. 
The significantly different time scales of the release of $\alpha$-elements
and iron to the interstellar medium results in a time delay of the order
of $\sim 1$\,Gyr in the \feii\ enrichment.
Detecting \feii\ emission at high redshift comparable to the relative strength
observed in quasars at lower redshift indicates that the formation of the
stars which had released the iron had occurred at least $\sim 1$\,Gyr earlier.
Therefore, the line ratio of $\alpha $-element vs. iron emission can be used 
as a cosmological clock.

Given the complexity of the \feii\ emission spectrum accurate iron abundances
are not yet easy to deduce and the full synthesis of the individual AGN spectra
is required \citep{Veetal99} which seriously complicates such measurements.
However, the UV \feii\ multiplets around \mgii\ ($\lambda \lambda \sim 
2000 - 3000$\,\AA ), the most common strong iron multiplets among AGNs 
\citep{Wietal80,Gran81,Neetal85}, are probably the most promising for 
obtaining at least some information on the relative iron abundance 
\citep[e.g.,][]{WaOk67,WNW85,HaFe99}.
A suitable spectral feature for testing the presence of $\alpha$-elements 
is the \mgii\ resonance doublet at $\lambda \lambda 2795, 2803$\,\AA\ 
(hereafter \mgii $\lambda 2798$). 
Since \mgii\ and \feii\ have similar ionization potentials and are expected 
to originate in the same partially ionized zone of the excited gas, the ratio 
of these lines provides a valuable measure to estimate the $\alpha $-element 
vs. iron abundance ratio, at least on a relative scale \citep{HaFe99}. 
Observations of the \mgii $\lambda 2798$ and of the rest-frame UV-multiplets 
of \feii\ in the spectra of several high-redshift quasars have been published 
by \citet{Hietal93}, \citet{Eletal94}, \citet{Kaetal96}, \citet{Taetal97}, 
\citet{Yoetal98}, \citet{Muetal98}, \citet{Muetal99},
\citet{Thetal99}, \citet{Kuetal01}, and \citet{Gretal01}. 
Generally, fairly strong \feii\ emission was found, indicating a very early 
cosmic epoch of the first star formation in the high-z quasar host galaxies. 

In order to derive more accurate and more quantitative information on the 
first star formation epoch in such objects we studied a small sample of six 
quasars with redshifts $z \simeq 3.4$ using uniformly processed and analyzed 
high-quality near infrared spectra covering (in the observer's frame) the 
wavelength range $(1.0 \ltsim \lambda \ltsim 2.5 \mu$m).
The quasars were selected to be accessible at the time of observation and
to be bright enough to be observed within reasonable integration times with 
ESO's 3.5\,m NTT.
Assuming H$_o  = 72$ km\,s$^{-1}$ Mpc$^{-1}$, $\Omega _M = 0.3$, 
$\Omega _\Lambda = 0.7$ \citep{CPT92,Caetal99,Peetal99,Freetal01} this 
redshift corresponds to a cosmic epoch of $\sim 1.8$\,Gyrs, i.e. about 
10\%\ of the current cosmic age. For the selected redshifts the strong 
\mgii $\lambda 2798$ emission line and most of the strong ultraviolet \feii\ 
emission at $\lambda \lambda 2300 - 2600$\,\AA\ are redshifted into the NIR 
J-band, while the \Hbeta\ and  [\oiii ]$\lambda \lambda $4959,5007 lines 
together with the optical \feii\ emission are shifted to the K-band wavelength
region.

We covered practically the whole restframe wavelength range 
$\lambda \lambda  2100 - 5600$\AA\ continuously. This made it possible to 
derive reliable continuum fits and, thus, at least internally rather accurate
flux values for the ultraviolet and optical \feii\ emission. 

\section{Observations and Data Analysis}
The coordinates, redshifts, and the apparent brightness of the six quasars are
listed in Table 1. All observations were carried out on Oct.\,19 -- 22, 1999 
using the SofI (Son of ISAAC) NIR spectrometer and camera attached to the 
3.5\,m NTT telescope of the European Southern Observatory at La Silla, Chile. 
The seeing was typically $\approx 1$\arcsec\ during the first three nights 
and $\approx 1$\arcsecpoint 7 during the fourth night.
The detector was a 1024 $\times$ 1024 Hawaii HgCdTe array. All spectra were 
recorded in the longslit mode (1\arcsec $\times$ 290\arcsec ). The pixel scale
was 0\arcsecpoint267/pixel. To obtain the entire near 
infrared spectral range from $\sim$0.95$\mu$m to $\sim$2.5$\mu$m we used the 
blue grism ($0.95 - 1.64\mu$m) and the red grism ($1.52 - 2.52\mu$m) in 
first order. 
This setting provides a reasonable spectral overlap in the H-band region
($\approx 0.12 \mu$m).
For flux calibration and correction of the strong atmospheric absorption 
features we observed the standard stars listed in Table 1 
\citep{Eletal82,Peetal97} several times during each night.

The observing epochs and exposure times for the quasars and standard stars
are listed in Table 2.
Making use of the SofI long-slit mode we changed the 
location of the object along the slit by 60\arcsec\ (with an additional 
random offset within 5\arcsec ) for subsequent exposures to optimize the
sky-correction.

The dark frames were recorded with the same exposures time as the science 
frames. The average dark frames of a specific exposure time were spatially 
extracted and the 1\,D countrate distributions were fitted with a low-order 
polynomial. The low-order polynomial fits were subtracted to correct for
the dark current signal.

For wavelength calibration we took xenon-neon comparison spectra in the blue 
and red setting of the gratings in lamp on and lamp off mode.
The latter measures the IR background contamination. Its subtraction
allows access to weaker spectral lines.  
Because the location of the spectral lines were reproduced within less than
0.3 pixels, we used a single XeNe-spectrum for the wavelength 
calibration for all three nights.
In the blue wavelength range ($0.95 - 1.64\mu$m) the 2\,D wavelength
calibration based on 19 spectral lines yielded a pixel scale of 
6.93 \AA /pixel and a wavelength mean error of 0.50\,\AA . 
For the red wavelength range ($1.52 - 2.52\mu$m) the corresponding 
values (based on 16 spectral lines) are 10.19 \AA /pixel and 
0.83\,\AA . The FWHM spectral resolution measured using strong night sky 
emission lines was about R\,$\simeq 600$ (blue) and R\,$\simeq 700$ (red). 

Flatfield frames taken with the internal lamps turned out not to be usable 
because of offsets (of up to 4 columns in the spatial direction).
Furthermore, normalized dome flatfield exposures of the same night 
introduced additional structures on large scales.
Therefore, we used the science exposures to compute a sky flatfield 
for the flatfield correction. 
Since, due to our observing strategy, the location of an object was 
different for each individual exposure, we computed the median of the 
normalized science frames. These median frames were used to compute a 
normalized sky flatfield according to the procedure outlined by \citet{Horn86}.
These flatfield frames were used to apply the 2\,D flatfield correction to 
the science frames.

We derived and subtracted the night sky intensity for each frame individually 
because the night sky intensity showed significant variations on time scales 
of 180\,sec - 240\,sec. For this purpose a 3$^{rd}$ order polynomial fit was 
calculated for each wavelength element to fit the spatial intensity 
distribution of the night sky emission. 
The fit was based on 22\arcsec\ wide regions which were at least 5\arcsec\
separated from the targets, quasar and standard star, both of which were
point source like.
The remaining residua after the sky correction were removed by subtracting
the object-free part of a sky-corrected frame containing the same residua.

The observed standard stars were used to correct the strong atmospheric 
absorption bands separating the J- and H-band and the H- and K-band, 
respectively, and other atmospheric (absorption) features. 
The range of the atmospheric transmission properties during the observing 
run is shown in Fig.\,1. 
  
To obtain a sensitivity function to transform the observed count rates to 
flux units we assumed that the IR spectral energy distribution of a A-type 
star and G-type star can be described with a black body energy distribution 
of a specific T$_{eff}$. 
We used the black body temperatures given by \citet{Kuru92} for the 
different spectral types of the stars we observed. 
We applied T$_{eff}$=5850\,K (HD 25402, HD 210395),
T$_{eff}$=8250\,K (HD 19904, HD 205772), and T$_{eff}$=9750\,K (HD 38921)
to compute a black body energy distribution for the observed wavelength 
range. 
Next, these black body spectra were scaled to match the apparent magnitudes of 
the observed standard stars in the J-,\,H-, and K-band.
For each standard star a sensitivity curve was calculated. 
For HD 25402 and HD 210395 we had to
calculate the apparent near infrared magnitudes based on the apparent V-band 
magnitude while for HD 19904, HD 38921, and HD 205772 we could use the
J,H, and K magnitudes given by \citet{Eletal82}.
In general, the sensitivity functions given for these two subsets are 
identical within 
less than 4\% . But they differ by $\sim 10$\%\ at the long wavelength end 
of the blue and red wavelength range. 
Hence, we used the J,H,K-band based sensitivity functions provided by 
HD 19904, HD 38921, and HD 205772 to derive a mean sensitivity function 
for the blue and red wavelength ranges. 

The quasar spectra were extracted using the \citet{Horn86} extraction 
routine.
The width of the spatial profile for the quasar spectra was the same as 
measured for the stars. Hence, the quasar spectra were treated as point 
sources. Since some individual spectra of low quality where eliminated
during the coaddition procedure, the effective integration time 
for the quasars is somewhat lower than the sum of the individual
exposure times (Tables 2,3).

The individual spectra of the quasars were corrected for atmospheric
absorption using appropriately scaled transmission functions provided 
by observed spectra of the standard stars.

To correct for cosmic-ray events the individual 1\,D spectra were compared 
with one another.
We calculated for each wavelength element the mean and standard deviation
among the available 1-D spectra, excluding the smallest and largest flux 
measurement. If these excluded measurements deviate by more than three times
the standard deviation, they were replaced by the calculated mean 
value, respectively.
For each quasar weighted mean spectra were calculated. The weight was given
by the mean signal-to-noise ratio in the continuum across the spectrum.
Finally, we transformed the quasar spectra to those emitted in the quasar
restframe. Following \citet{Pete97} the flux conversion was carried out 
according to

$$
F'_{obs}(\lambda _o) = F_{rest}(\lambda _1) / (1 + z)^3
$$

\noindent
The resulting restframe quasar spectra are displayed in Fig.\,2. 


\section{Modelling of the Quasar Spectra}

While at least the narrow emission component of the strong and relatively 
isolated \mgii\ $\lambda 2798$ doublet
appears to be measured easily in QSO spectra, the \feii\ emission forms 
broad emission blends caused by the superposition of several ten thousand 
discrete lines and the intrinsically large quasar emission line widths
\citep[e.g., Wills et al. 1985, hereafter WNW85;][]{Veetal99}. 
Due to the large number of merging lines and the difficulty of deriving a 
local continuum level, it is not possible to measure these lines individually.
However, as suggested and demonstrated by WNW85, it 
is possible to derive the \feii\ emission strength by decomposing the quasar 
spectrum into several well defined components. Therefore, we assumed our 
observed spectra to consist of a superposition of the 
following four components, 
    (i)   a power law continuum (F$_\nu \sim \nu ^{\alpha}$),
    (ii)  Balmer continuum emission,
    (iii) a pseudo-continuum due to merging \feii\ emission blends,
and (iv)  an emission spectrum of other individual broad emission lines.

\subsection{Non-stellar Continuum}
We first determined component (i), the underlying non-stellar power law 
continuum, from spectral windows which are free (or almost free) of 
contributions by the components ii to iv. For this purpose it was
very important to have spectra covering the (rest frame) optical region 
up to $\sim 5500$ \AA\ with one of the continuum windows at $\sim 5100$ \AA ,
which is nearly free of emission line contamination and hydrogen continua. 
Only minor \feii -emission can be expected for this wavelength region 
\citep[WNW85;][]{Veetal99} and so will not significantly affect the
continuum setting.
For three of our quasars (Q 0103-260, Q 0256-0000, and Q 0302-0019) we 
also have restframe UV spectra available,
covering \ovi $\lambda 1034$ to \ciii ]$\lambda 1909$ \citep{DiHa01}
with continuum windows at $\lambda \lambda \approx 1330 - 1380$\,\AA\ and 
$\lambda \lambda \approx 1440 -1470$\,\AA\ which are nearly uncontaminated by 
line emission as well \citep{Veetal99}.
For those quasars where we had no access to the short wavelength UV part
of the spectrum we used continuum windows at $\sim 2100$\,\AA\ to 
estimate the continuum strength. 
The uncertainty introduced by estimating the continuum based on at least
two of the spectral ranges described above, is estimated to be of the 
order of $\sim 10$\,\% .

The UV data allow us even better constraints to be placed on the non-stellar 
continuum for these objects.

\subsection{Balmer Continuum}
The Balmer continuum emission was modeled according to the following
procedure: Assuming case B conditions the flux ratio of the integrated Balmer 
continuum emission (BaC) and of H$\beta \,\lambda$4861 is given by  
I(BaC)/I(H$\beta$) = 3.95 T$_4^{0.4}$, with T in units of $10^4$\,K 
(e.g., WNW85).
But for significant optical depth in the Balmer emission lines and especially 
in the Balmer continuum, large deviations from this simple relation are 
expected.
Optically thick Balmer continuum emission can be described by a blackbody 
spectrum at all wavelengths \cite[e.g.,][]{MaSa82}. But since the 
absorption cross section decreases as $(\nu / \nu_{BE})^3$, the continuum
emission will become optically thin at some wavelength. To get an estimate of 
the Balmer continuum emission spectrum for the purpose of model fitting we 
assumed clouds of uniform temperature (T$_e = 15000$\,K) which are partially 
optically thick. In this case the Balmer continuum spectrum can be described 
by

$$ F_\nu ^{BaC} = F{\rm '}_\nu ^{BE} B_\nu (T_e) (1 - e^{-\tau _\nu}); 
   \quad \nu \geq \nu_{BE}$$

\noindent
with B$_\nu$(T$_e$) as the Planck function at the electron temperature T$_e$
\citep{Gran82}.
$\tau _\nu$ is the optical depth at the frequency $\nu$. $\tau _\nu$ can be
calculated in terms of the optical depth $\tau _{BE}$ at the Balmer edge
using 

$$ \tau _\nu = \tau _{BE} ({\nu \over \nu_{BE}})^{-3}$$

\noindent
and F${\rm '}_\nu ^{BE}$ is a normalized estimate for the Balmer continuum 
flux density at the Balmer edge at $\lambda = 3646$\,\AA .\\
After subtraction of the power-law continuum component, the strength of the 
Balmer continuum emission can be estimated from the flux density 
at $\lambda \simeq 3675$\,\AA\ , since at this wavelength there is no 
significant contamination by \feii\ emission \citep[WNW85;][]{Veetal99}.  
The $\lambda 3675$\,\AA\ restframe flux density level was therefore 
used to normalize the Balmer continuum spectrum.
At wavelengths $\lambda \geq 3646$\,\AA\ higher order Balmer lines are
merging to a pseudo-continuum, yielding a smooth rise to the Balmer edge
(WNW85).
We used the results of the model calculations provided by \citet{StHu95} 
(case B, T$_e = 15000$\,K, n$_e = 10^8 - 10^{10}$\,cm$^{-3}$).
We calculated several Balmer continuum spectra for T$_e = 15000$\,K and
$0.1 \leq \tau _\nu \leq 2$ to obtain Balmer continuum template spectra.
These Balmer continuum templates were supplemented for $\lambda > 3646$\AA\ 
with high order Balmer emission lines with 
$10 \leq n \leq 50$, i.e.\, H$\vartheta $ and higher.

\subsection{\feii\ Emission}
Calculating the \feii\ emission spectrum, is much more difficult and the 
influence of unknown parameters such as metalicity, pumping by the incident 
continua, line fluorescence, emission line transport and turbulence 
velocities, which affect the emergent \feii\ spectrum,
are still not well understood 
\citep[e.g., WNW85;][]{Neetal85,BaPr98,SiPr98,Veetal99,CoJo00}. However, 
in spite of these 
uncertainties WNW85, \citet{Laetal97}, and \citet{Mcetal99}
and others have shown that the \feii\ emission spectrum of Seyfert\,1 
galaxies and quasars can be modelled.
Therefore, we fitted the \feii\ emission in our quasar spectra using scaled 
and broadened empirical \feii\ emission template spectra to derive relative 
\feii\ emission strength values.
For the ultraviolet wavelength range these templates had been extracted
from HST observations of I\,Zw1 by \citet{VeWi01}. The
optical \feii -emission template, extracted from ground-based spectra of
I\,Zw1, was kindly provided by T.\,Boroson.

\subsection{Strong Broad Emission Lines}
The broad emission lines of \mgii $\lambda 2798$ and of \Hbeta $\lambda 4861$ 
were fitted in our spectra with Gaussian components to measure the
integrated line flux. Generally, the \mgii $\lambda 2798$ emission line 
profile could be reconstructed with two Gaussian components, one narrow and 
one broad and blueshifted. 
For the \Hbeta $\lambda 4861$ emission line profile we used the 
same approach. While the width of the narrow components in both lines tend to 
be of the same order (FWHM(\mgii ) = 2850$\pm$460 km\,s$^{-1}$ and
FWHM(\Hbeta ) = 2950$\pm$360 km\,s$^{-1}$), the broad component which we used
for \Hbeta\ was significantly broader than for \mgii\ (FWHM(\Hbeta ) = 
10000$\pm$1000 km\,s$^{-1}$ and FWHM(\mgii ) =6200$\pm$700 km\,s$^{-1}$).
However, the approach to reconstruct the \mgii\ and \Hbeta\ emission line 
profiles with two Gaussian components was only chosen to measure the line 
flux; each individual component has no physical meaning by itself.

\subsection{Internal Reddening}
The effect of internal reddening on broad-emission lines is still unclear.
In recent years growing evidence has appeared for the presence of large 
amounts of dust (M$_{dust} \gtsim 10^8$\,M$_\odot$) in the host galaxies 
of high redshift quasars 
\citep[e.g.,][]{Guetal97,Guetal99,Caetal00,Ometal01}. It is 
assumed that the dust is distributed in a kiloparsec-scale warped disk 
\citep{Saetal89} which is illuminated by the central AGN. The observed 
dust emission spectra from 3 to 30 $\mu$m can be explained by such a 
model as shown by \citet{Anetal99} and \citet{Wietal00}.
However, since the present spectra are typical quasar spectra with 
prominent broad emission lines, it is unlikely that the Broad-Line Region
is significantly blocked by dust \citep[e.g.,][]{NeLa93}.
Even if the radiation has to pass through an (external) dust screen the 
extinction of the \mgii\ and the UV \feii -emission (having about the 
same mean wavelength) will be comparable and the line ratio is not
expected to be significantly modified. 
However, in the case of dust located within the line emitting gas, the 
situation is more complicated. Since \mgii $\lambda 2798$ has a larger 
optical depth than that of the \feii -emission, the \mgii\ line 
emission will suffer more resonance scattering, resulting in a longer 
effective internal lightpath, and thus be more weakened than the \feii 
-emission. This would result in a larger \feii /\mgii\ ratio. To investigate 
the influence of internal dust reddening, detailed model calculations are 
required. Such calculations are beyond the scope of the current paper;
however this effect will be investigated for an upcoming study of the 
evolution of the \feii /\mgii\ ratio for quasars at redshifts of $0 < z < 5$.


\section{Results}

The results of our analysis are given in Figs. 3 -- 5 and in Table 4. 
We fitted five spectral wavelength ranges ($\lambda \lambda $
$\sim 2200-2750$\,\AA , $\sim 2860-3080$\,\AA , $\sim 3400-3650$\,\AA , 
$\sim 4500-4650$\,\AA , $\sim 5050-5400$\,\AA ) in the quasar spectra.
These windows were selected since they correspond to the spectral range 
of our template spectra and are not too strongly affected by atmospheric 
absorption in our IR-spectra.
The strength of the Fe-emission templates was varied.
We also varied the strength and the optical depth $\tau _\nu$ of the Balmer 
continuum emission as indicated above.
We calculated a minimal $\chi^2$ to determine the best fit. 
In order to estimate the accuracy of our fits, we calculated fits also 
for a sample of different parameters around the best fit values and we 
determined $\chi ^2$ as a function of the parameter values.
From the widths of the resulting $\chi ^2$ distributions we then 
calculated mean errors for these parameters.

To compare the spectal properties of the high-z quasars with the quasar 
population at lower redshift, we calculated a local mean quasar spectrum.
This quasar spectrum consists of 101 quasars which were taken from a large 
quasar sample \citep{DiHa01}. The quasars were selected to have 
$z \leq 2$ and to cover a luminosity range comparable to the $z \simeq 3.4$
quasars under study. The criterion of comparable luminosity minimizes
luminosity effects on the emission line strength like the Baldwin effect 
\citep[e.g.,][]{OsSh99}. The average luminosity of the quasars which 
contribute to the local mean quasar spectrum amounts to 
log\,$\lambda$\,L$_\lambda (1450$\AA $) = 43.5 \pm 0.3$ erg\,s$^{-1}$. 
The average luminosity of the six high-z quasars is 
log\,$\lambda$\,L$_\lambda (1450$\AA $) = 43.9 \pm 0.2$ erg\,s$^{-1}$.

In Fig. 3 and 4 we present the multicomponent fit for the six high-z quasar
spectra and in Fig.\,5 for the local mean quasar spectrum.
The individual components are also shown together with the fit. 
In addition, the resulting residuum spectra are displayed, too.
Although the observed quasar spectra are well represented by the multicomponent
models, several broad emission features can be seen in the residua.
The broad residuum in the wavelength range $\lambda \lambda \simeq 3800 - 4000$
\AA\ is caused by the Balmer emission lines \Hdelta\ up to H$\eta $,
because the BaC emission template we used includes Balmer emission lines
with $10 \leq n \leq 50$.
The second strong residual emission at $\lambda \lambda \simeq 3200$\AA\ 
is associated with the \feii -emission blends M6, M7 which are not included 
in the ultraviolet \fe -emission template we used.
In the residuum spectra of Q 0103-260, Q 0302-0019, and of the local mean
quasar spectrum some emission is detected for $\lambda \ltsim 2200$\AA .
This additional emission can be ascribed to \feii -emission which is not 
contained in our \fe -emission template but can be expected on the basis of
model calculations \citep[]{Veetal99}.
  
In addition to the relative flux of the \fe -emission blends we list in 
Table 4 also flux values for some other permitted and forbidden emission 
lines of interest in the observed wavelength interval.
The flux is given in units of the corresponding \mgii\,$\lambda 2798$ flux 
(i.e. the table lists in all cases the ratio between the observed emission 
flux and the flux of the \mgii\,$\lambda 2798$
line observed in the same spectrum).
The absolute \mgii\ restframe intensity is given at the bottom of Table~4.

For the \feii\ emission we list the integrated flux of most of the UV and
optical blends and multiplets, using the following designations
\citep[e.g.,][]{Phil78,Wietal80}
:
\feii 4570: multiplets 37,38,43 (4250 -- 4770 \AA ),
\feii 4924,5018: multiplet 42 (4800 -- 5085 \AA ),
\feii 5190: multiplets 42,48,49,55 (5085 -- 5500 \AA ),
\feii \,opt total: \feii\ flux 4250 -- 5500 \AA ,
\feii 2080: multiplets UV 83,91,93,94 + \feiii UV48 (2030 -- 2130 \AA ),
\feii 2500: multiplets UV 1,3,4,5,35,36,64  (2240 -- 2660 \AA ),
\feii 2680: multiplets UV 200,235,263,283 (2660 -- 2790 \AA ),                
\feii 2900: multiplets UV 60,78,277,215,231,255 (2790 -- 3030 \AA ),
\feii \,UV: total \feii\ flux 2200 -- 3090 \AA . 
The multiplets listed are the strongest ones expected in each range.
Note that, since the total flux values for entire blends or sums of 
multiplets often contain additional weak lines, these values are somewhat 
larger than the sums of the individual prominent blends.

According to Table 4 the mean emission-line ratio of 
I(\Hbeta\,$\lambda 4861$) / I(\mgii\,$\lambda 2798$) for the individual 
quasars is $1.14 \pm 0.16$
which, within the uncertainties, is consistent with the relative line 
strength of \Hbeta\ reported by WNW85 ($0.82 \pm 0.25$). 
The greater relative strength of \Hbeta $\lambda 4861$, measured here, 
might be caused by including a very broad component to measure the line 
strength. The outer
wings of the broad Gaussian component, which we used to determine I(\Hbeta ),
contributes up to $\sim 20$\,\%\ of the total I(\Hbeta ). 

The strength of the Balmer continuum emission (BaC) in units of 
I(\mgii $\lambda 2798$) as listed in Table 4 shows some variation between 
the six high-z quasars.
In particular, Q 2227-3928 shows quite weak BaC emission.
The relative BaC strength without Q 2227-3928 amounts to
I(BaC)/I(\mgii ) = $5.20 \pm 0.38$. Including Q 2227-3928, the value
drops to I(BaC)/I(\mgii ) = $4.57 \pm 0.71$. However, within the errors
the relative BaC emission strength is consistent with the strength we 
measured for the local mean quasar spectrum, 
I(BaC)/I(\mgii )$_{local}$ = $5.02 \pm 0.51$.

Whether the ratio of the optical-range \feii -emission of the high-z
quasars relative to \mgii\ 
differ from the \feii /\mgii\ ratio of the local quasar population
could not be studied because our local mean quasar spectrum does not cover
the optical \feii\ emission at $\lambda > 4200$\,\AA .
The average relative strength of the optical \feii -emission of the six
high-z quasars was found to be I(\feii \,opt)/I(\mgii ) = $1.50 \pm 0.16$. 
This value is in good aggreement with the ratio $1.21 \pm 0.73$ given by 
WNW85, who studied quasars with $0.12 \leq z \leq 0.63$. 
This comparison with the WNW\,85 \feii /\mgii\ ratio for local quasars
thus indicates no significant difference, i.e. no evolution of the 
\feii /\mgii\ ratio up to $z \simeq 3.4$.

More interesting is the integral strength of the ultraviolet \fe -emission 
($2200 - 3090$\,\AA ) relative to \mgii $\lambda 2798$. 
We obtain for the six high-z quasars an average of
I(\feii \,UV)/I(\mgii ) = $3.72 \pm 0.20$. 
Within the uncertainties there is no difference compared with 
I(\feii \,UV)/I(\mgii ) = $3.82 \pm 0.40$ derived from the local mean 
quasar spectrum (Tab.\,4).

We anticipate that the approach adopted in this work is relatively more 
accurate than those adopted in the studies by \citet{Thetal99}, 
\citet{Kaetal96}, and \citet{Muetal99}. 
This is partly due to our use of data with a large, mostly continuous spectral 
coverage, which allow a much better determination of the underlying continuum.
Also, a better account was made of the Balmer continuum emission through a 
direct fitting thereof; thanks again to the long and continuous spectral 
coverage. Finally, we use {\em empirical} UV \feii\ emission 
templates \citep{BoGr92,VeWi01}
as opposed to the theoretical templates from the study by WNW85. 
As discussed by \citet{Thetal99} the latter approach includes 
several assumptions on the relative strengths of the emission lines and is 
clearly an approximation.
Although, the relative line strengths, I(\feii \,UV)/I(\mgii ), measured
here are somewhat lower than the Thompson et al., Kawara et al., and the 
Murayama et al. quoted values, they are not in immediate contradiction, given
the larger uncertainties in the earlier measured values.
It is interesting to note that our results are much closer to the 
I(\feii \,UV)/I(\mgii ) values predicted by the theoretical models, 
applied to observed AGN spectra, \citet{NeWi83} and 
WNW\,85.

By combining the above result for the UV and optical FeII emission we obtain 
for our high-z quasars 
I(\feii \,UV + \feii \,opt)/I(\mgii ) = $5.22 \pm 0.29$. This value is 
lower than most numbers quoted in the literature, but again close to the 
I(\feii $_{tot}$)/I(\mgii ) value predicted by theoretical models 
\citep[WNW85]{NeWi83}.
 
The I(\feii \,UV)/I(\mgii ) ratio which we find for our local quasar sample
is comparable to the corresponding value reported for the mean quasar 
spectrum of the LBQS sample (containing quasars with redshifts in a range of 
$z\simeq 1-2$) with I(\feii \,UV)/I(\mgii ) $\simeq 3.7$ \citep{Fraetal91}
but below the ratios given by \citet{Thetal99} (4.3 -- 5.3) 
and \citet{Muetal99} (8.9).
As discussed above, this difference is likely due to our better observational
data and the more accurate approach adopted here.

\section{Discussion and Conclusion}

As pointed out in Section 4 our accurate derivation of the
relative \feii -emission strength for a sample of six high-z quasars,
give lower or much lower 
I(\feii \,UV)/I(\mgii ) ratios than the values reported
in the literature for intermediate and high
redshift quasars \citep[e.g., WNW85;][]{Thetal99,Muetal99}. 
Since the earlier results are consistent with our data to within their higher 
error limits, we assume that the differences result from our improved 
multi-component
fit of the observed spectra, which was facilitated by the wide
wavelength range and accurate calibration of our observational data.
On the other hand, the I(\feii \,UV)/I(\mgii ) ratios for our $z\simeq 3.4$ 
quasars reported in this paper are comparable to the corresponding mean ratio 
for low-z quasars, which we determined from the local mean quasar spectrum
\citep{DiHa01}.
As our high-z and local quasars do not differ significantly in any other 
spectral property, it appears unlikely that the strong \feii\ emission seen 
also at $z = 3.4$ can be explained by any other mechanism than comparable 
relative abundance of Fe in our $z\simeq 3.4$ quasars. 
Since, as pointed out in Section 1, 
according to the present chemical evolution theory of galaxies 
Fe is produced mainly by Type Ia SNe which begin to explode about
$\sim 1$ Gyr after the beginning of star formation, the observed high Fe
content of the broad line region (BLR) gas of our quasars confirms that the 
star formation in the host galaxies of high-z quasars started at a very 
early epoch. 

An evolutionary model of the Fe enrichment in a galaxy following the 
initial 
starburst has been calculated by \citet{Yoetal98}. Although, as pointed 
out by Yoshii et al., evidence that the initial mass function may have been 
different for the first stars make such model calculations quantitatively
somewhat uncertain, it is clear from these computations that the
relative Fe abundance is rather low initially and starts to grow steeply 
about 1.0 Gyrs after the 
beginning of the star formation, reaching a maximum (at about 3 Gyrs
in the Yoshii et al. model), before
declining to the local value \citep[see also ][]{MaPa93,HaFe93}. 
The same temporal evolution of the Fe/Mg ratio is found for the giant 
elliptical galaxy model (M4a) presented by \citet{HaFe93}. Their model
predicts a strong increase of Fe/H at $\sim 1$\,Gyr after the beginning 
of the star formation, with the most rapid rise of the Mg/Fe ratio during 
the 1 - 1.38 Gyrs period.
Comparing our relative \feii\ emission strength with the model of 
\citet{Yoetal98} and \citet{HaFe93} we estimate an age of $\sim 1.5$\,Gyrs 
(with a formal error of $\pm 0.5$\,Gyrs) for the stellar population which 
produced the observed iron in our high-z quasars. (Because of 
our smaller \feii /\mgii\ emission ratio this value is slightly lower than the 
corresponding age derived by Yoshii et al. for B 1422+231 at $z = 3.6$). As 
discussed e.g. by \citet{Yoetal98} and \citet{Thetal99} a starburst age of 
this order and in 
this redshift range provides severe constraints on the allowed cosmological 
parameters as realistic cosmologies require a cosmic age
at the epoch of the light emission which is larger than the age of the star 
burst. Like the earlier results of \citet{Yoetal98} and \citet{Thetal99} our 
new data basically rule out cosmologies with $\Omega _M$=1. 
On the other hand, if
the cosmological parameters are known, it is possible to infer from the 
observed starburst age the redshift at which the star formation started.
Although there is still some uncertainty about the cosmological parameters, 
at present a universe with H$_o = 72$\,km\,s$^{-1}$\,Mpc$^{-1}$ 
\citep{Freetal01}, $\Omega _M$=0.3, $\Omega _\Lambda = 0.7$ 
appears to be a good approximation. 
For these parameters the age of the universe at the time when the light was 
emitted by our $z\simeq 3.4$ quasars was about 1.8 Gyrs. Hence, the star
formation age of $\sim 1.5$ Gyrs derived for our high-z quasars results in an 
epoch of the beginning of star formation in these objects of $\sim 0.3$ Gyrs, 
corresponding (with the above parameters)
to $z_f \gtsim 10$. 
The redshift $z_f$ can be reduced too if smaller values
for H$_o $ and/or $\Omega _M$ are assumed. On the other hand, for  
$\Omega _\Lambda = 0.7$, $\Omega _M = 0.3$ and H$_o > 95$\,km\,s
$^{-1}$\,Mpc$^{-1}$ the age at $z = 3.4$ becomes shorter than the timescale
for the production of the observed amount of Fe in SNe type Ia, which
seems to rule out such high values of H$_o$. 
However, we would like to note that the above age estimates are 
based on a relatively crude chemical evolution model and on the 
assumption that the SNe type Ia in the early universe have similar 
properties as those observed in the local universe. If, as suggested 
by various theoretical studies, SNe type Ia can, under certain 
conditions, also be produced by massive stars with a much shorter 
evolutionary time scale than that assumed in the standard scenarios, 
some SNe type Ia progenitors could have much shorter life times 
\citep[e.g.,][]{IbTu84,SmWy92}. It cannot be excluded that in the 
Population III of massive galaxies at early cosmic epochs short-lived 
SNe type Ia progenitors were more common than in the local universe. In 
this case the above estimates would provide upper limits only.
To constrain the cosmological parameters further it will be important
to extend this type of analysis to known QSOs with $z>5$. 

\acknowledgments{ 
This work has benefited from support of NASA grant NAG\,5-3234 and of the 
Deutsche Forschungsgemeinschaft, project SFB 328 and SFB 439 (MD).
MV acknowledges financial support from the Columbus Fellowship.

\newpage

\figcaption[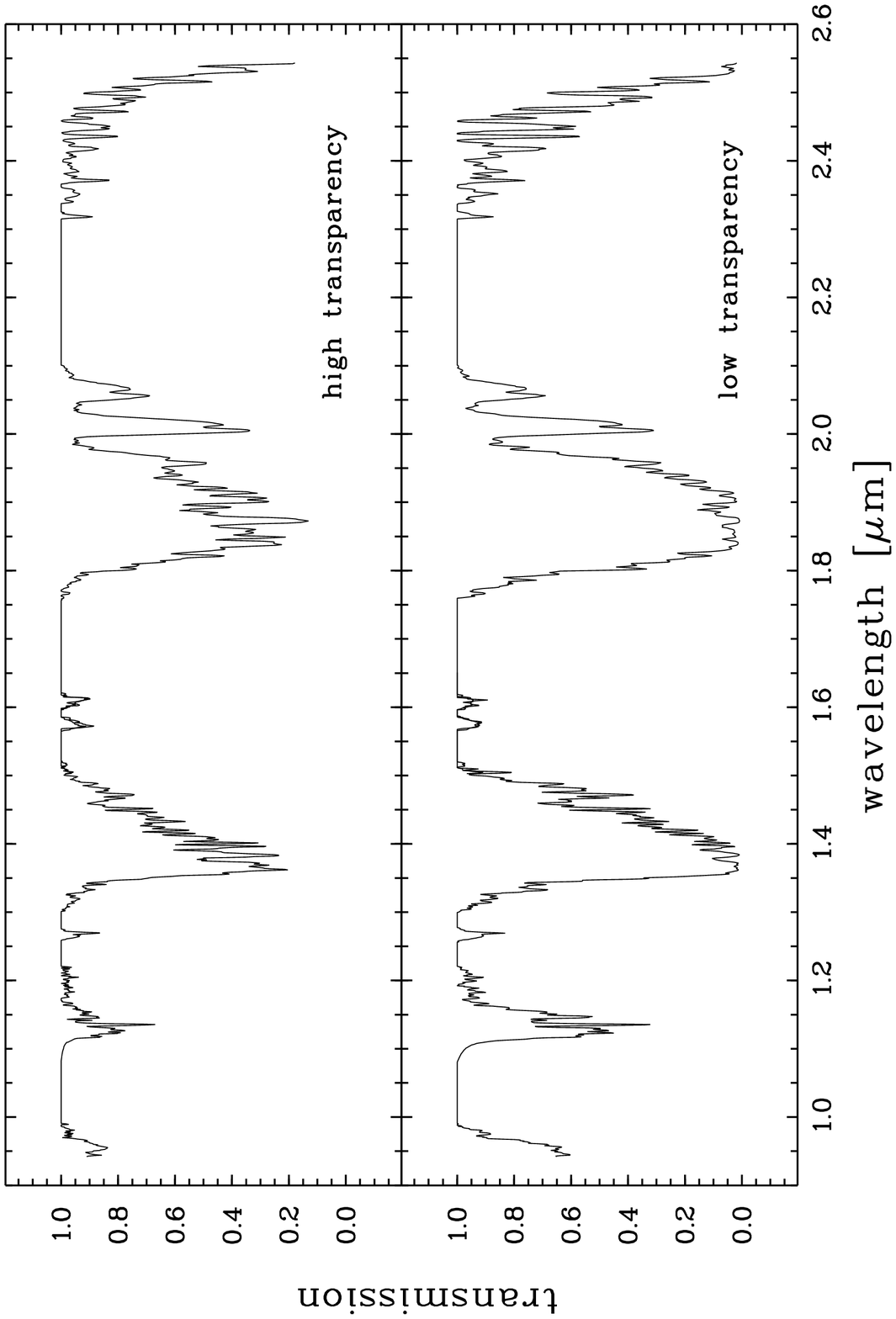]{The atmospheric transmission during the observing 
                   run obtained from the standard stars as a function of 
                   wavelength for optimal conditions (top panel) and poor 
                   conditions (bottom panel).}

\figcaption[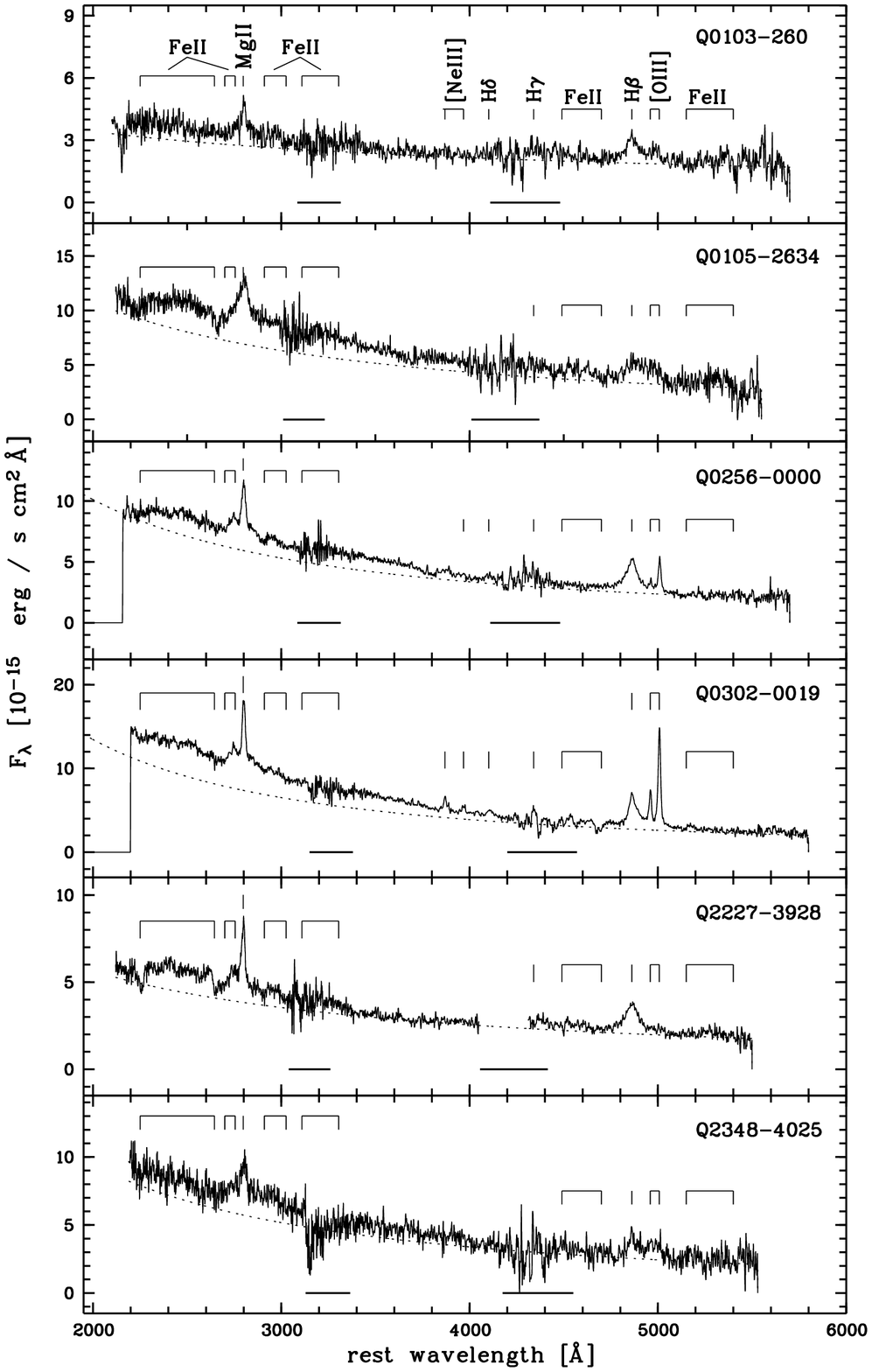]{The quasar spectra transformed to the restframe.
                   The flux density is given in units of 
                   $10^{-15}$ erg\,s$^{-1}$\,cm$^{-2}$\,\AA $^{-1}$. 
                   The location of the strong atmospheric absorption bands is 
                   indicated by the horizontal bars for each quasar.}

\figcaption[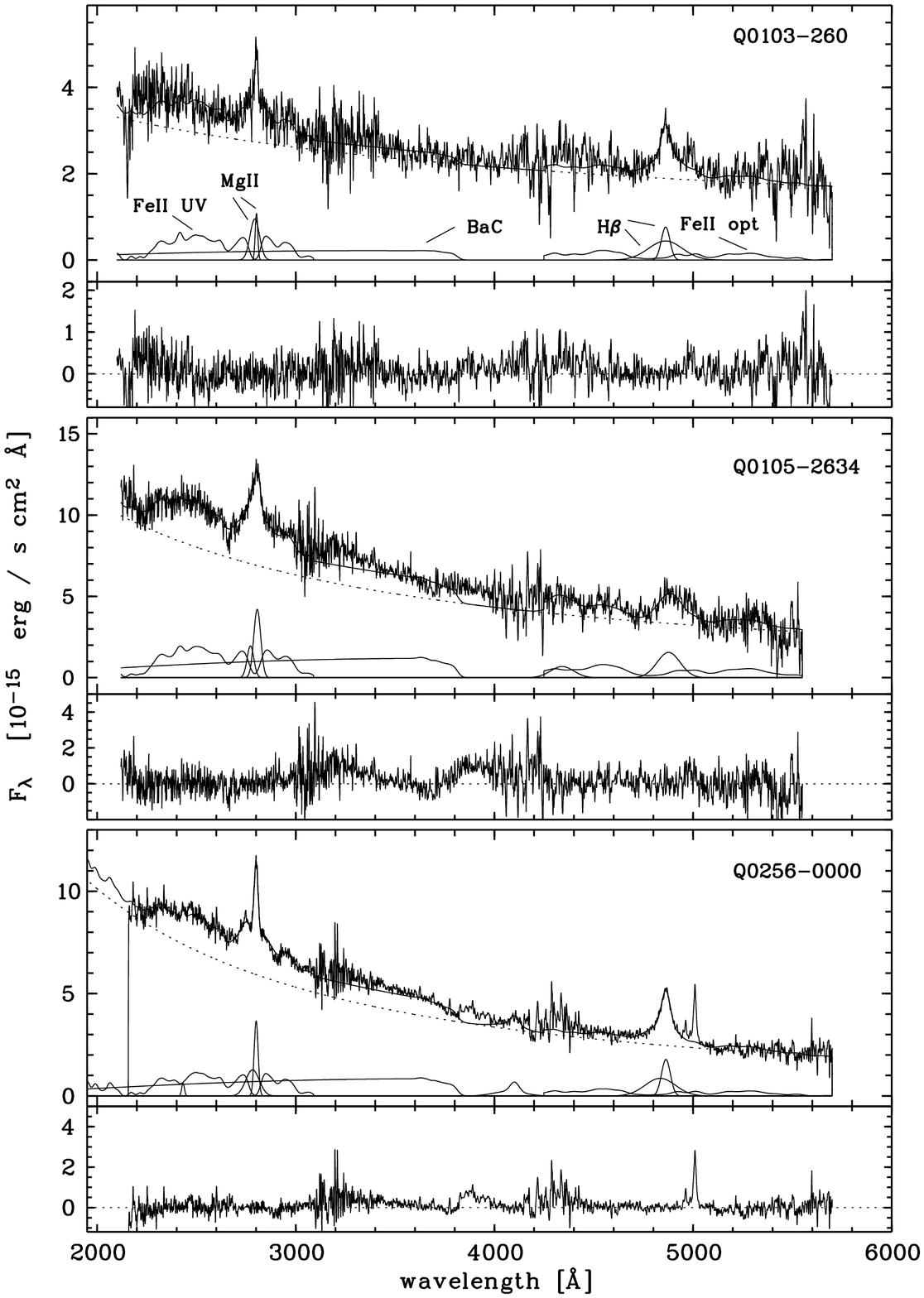]{The restframe quasar spectra together with results 
                    of the multi-component analysis are shown for 
                    Q\,0103-260, Q\,0105-2634, and Q\,0256-0000. 
                    In the top panel the quasar spectrum is shown together 
                    with the power law continuum fit (dotted line), the scaled
                    and broadened Fe-emission template, the scaled Balmer 
                    continuum emission, and the Gaussian components to fit
                    the \mgii $\lambda 2798$ and \Hbeta $\lambda 4861$ 
                    emission line profiles. The resulting fit is overplotted 
                    as a solid line.
                    In the bottom panel the quasar spectrum is shown after 
                    subtraction of these components.}

\figcaption[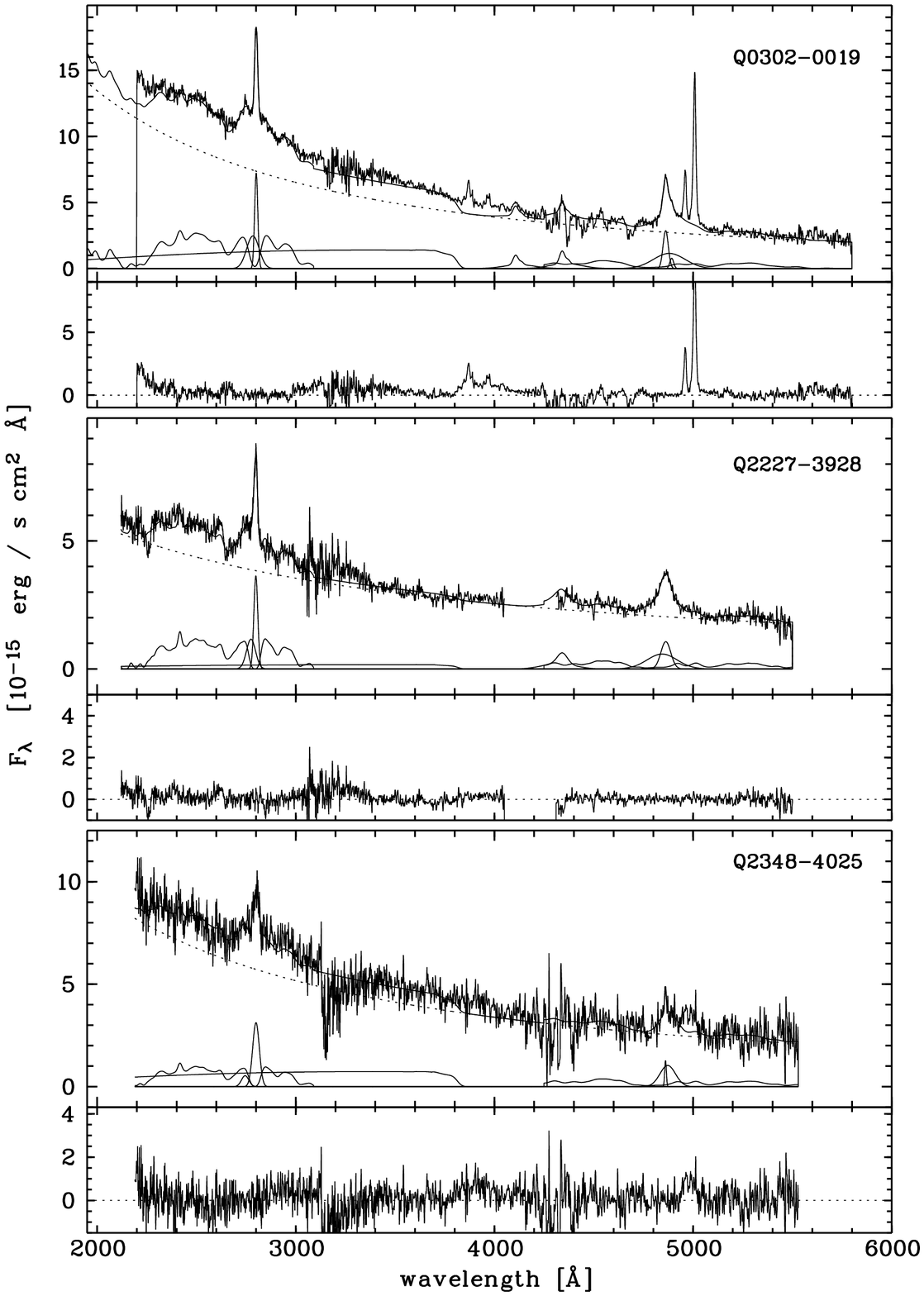] {Same as Fig. 3 for 
                    Q 0302-0019, Q 2227-3928, and Q 2348-4025.}

\figcaption[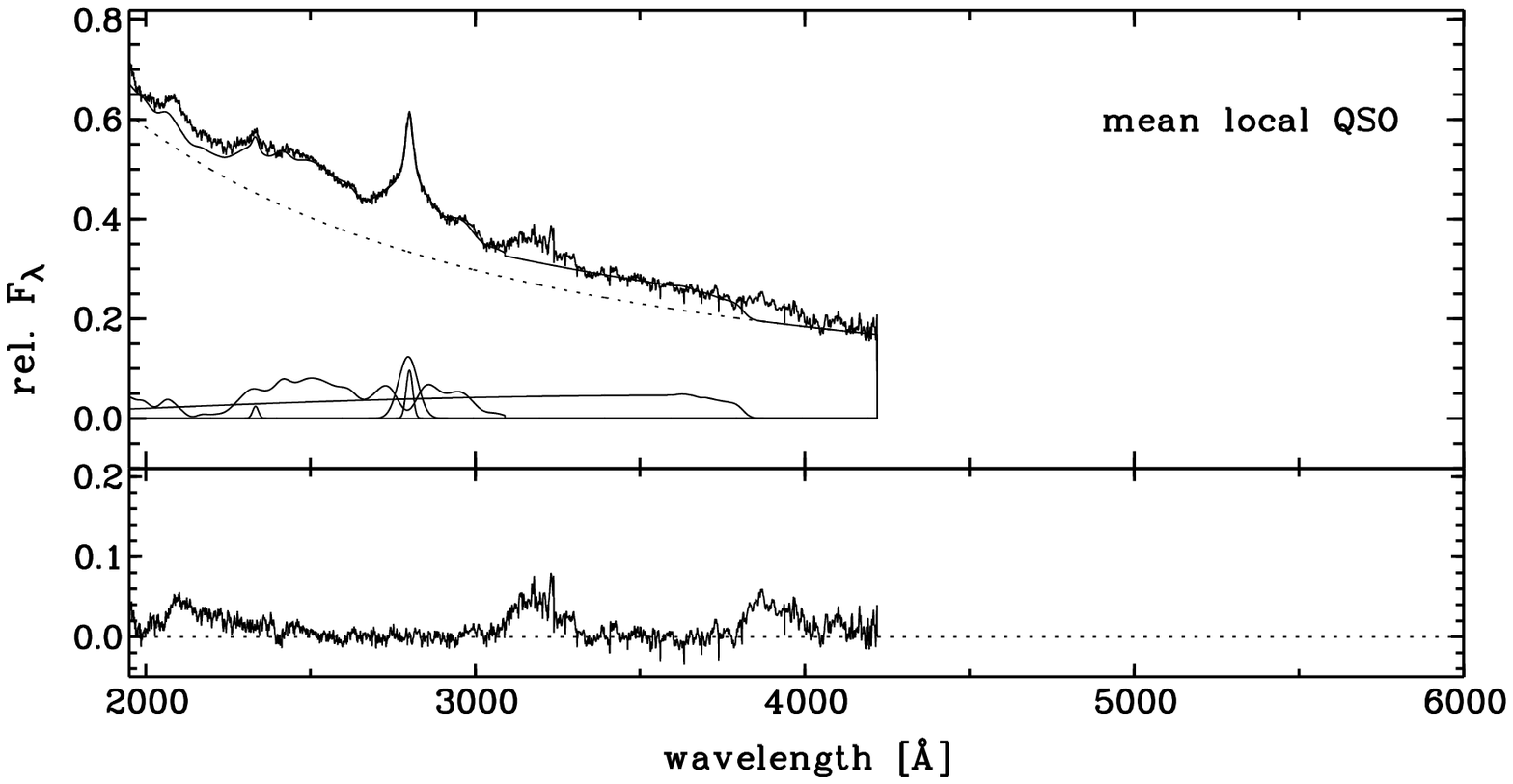] {Same as Fig. 3 for the local mean quasar
                    spectrum.}

\clearpage


\begin{figure}
\plotone{f1.eps}
\label{fig1}
\end{figure}

\begin{figure}
\plotone{f2.eps}
\label{fig2}
\end{figure}

\begin{figure}
\plotone{f3.eps}
\label{fig3}
\end{figure}

\begin{figure}
\plotone{f4.eps}
\label{fig4}
\end{figure}

\begin{figure}
\plotone{f5.eps}
\label{fig5}
\end{figure}

\clearpage


\begin{deluxetable}{lcccc}
\tablewidth{0pt}
\tablecaption{The quasars sample}
\tablehead{
\colhead{object} &
\colhead{RA(J2000.0)} &
\colhead{DEC(J2000.0)} &
\colhead{app.mag.\tablenotemark{a}} &
\colhead{z} \\
\colhead{(1)} &
\colhead{(2)} &
\colhead{(3)} &
\colhead{(4)} & 
\colhead{(5)} 
}
\startdata
quasar&$\,^h \,\,\, ^m \,\,\, ^s$&$\,\,\, \degr \quad \arcmin \quad \arcsec $&m$_r$&\\
\hline
Q 0103-260 &01 06 04.3&$-$25 46 53&18.82&3.375\\
Q 0105-2634&01 08 12.4&$-$26 18 20&17.30&3.488\\
Q 0256-0000&02 59 05.6&$+$00 11 22&17.50&3.377\\
Q 0302-0019&03 04 49.9&$-$00 08 13&17.60&3.286\\
Q 2227-3928&22 30 32.9&$-$39 13 07&18.60&3.438\\
Q 2348-4025&23 51 16.1&$-$40 08 36&18.10&3.310\\
\hline
star     & & &m$_K$ \tablenotemark{b} &type\\
\hline
HD 19904  &03 10 43  &$-$39 03 06&6.64 &A4III/IV\\
HD 25402  &04 00 32  &$-$41 44 54&8.36\tablenotemark{c} &G3V\\
HD 38921  &05 47 22  &$-$38 13 52&7.54 &A0V\\
HD 205772 &21 38 41  &$-$41 02 53&7.66 &A5IV/V\\
HD 210395 &22 11 02  &$-$39 32 51&7.98\tablenotemark{c} &G3V\\
\enddata
\tablenotetext{a}{\citet{VCV01}}
\tablenotetext{b}{\citet{Eletal82}}
\tablenotetext{c}{m$_v$}
\end{deluxetable}

\clearpage

\begin{deluxetable}{lcccccl}
\tablewidth{0pt}
\tablecaption{observation log}
\tablehead{
\colhead{object} &
\colhead{$\lambda \lambda$-range} &
\colhead{date} &
\colhead{t$_{int}$} &
\colhead{no.of exp.} &
\colhead{total$_{int}$}&
\colhead{comment}\\
\colhead{} &
\colhead{} &
\colhead{} &
\colhead{[sec]} &
\colhead{} &
\colhead{[sec]} &
\colhead{} \\
\colhead{(1)} &
\colhead{(2)} &
\colhead{(3)} &
\colhead{(4)} & 
\colhead{(5)} &
\colhead{(6)} &
\colhead{(7)}
}
\startdata
Q 0103-260 &blue&Oct.\,19/20&240&12&2880&seeing $\sim$0\arcsecpoint8\\
          &red &Oct.\,19/20&240&18&4320&seeing $\sim$0\arcsecpoint6\\
          &blue&Oct.\,20/21&240& 8&1920& \\
          &red &Oct.\,20/21&240&10&2400& \\
Q 0105-2634&blue&Oct.\,20/21&240&12&2880&seeing $\sim$1\arcsecpoint1\\
          &red &Oct.\,20/21&240& 6&1440&seeing $\sim$1\arcsecpoint2\\
Q 0256-0000&blue&Oct.\,19/20&240&12&2880&seeing $\sim$0\arcsecpoint7\\
          &red &Oct.\,20/21&240&12&2880&seeing $\sim$1\arcsecpoint2\\
          &blue&Oct.\,21/22&180&14&2520&seeing $\sim$2\arcsecpoint2, clouds\\
          &red &Oct.\,21/22&180& 3& 540&clouds\\
Q 0302-0019&blue&Oct.\,19/20&240&12&2880&seeing $\sim$0\arcsecpoint6\\
          &red &Oct.\,19/20&240&15&3600&seeing $\sim$0\arcsecpoint7\\
          &blue&Oct.\,20/21&240&12&2880&seeing $\sim$1\arcsecpoint0\\
          &red &Oct.\,20/21&240&12&2880&seeing $\sim$1\arcsecpoint0\\
Q 2227-3928&blue&Oct.\,19/20&180&16&2880&seeing $\sim$1\arcsec\\
          &red &Oct.\,19/20&180& 8&1440& \\
          &blue&Oct.\,20/21&240& 6&1440& \\
          &red &Oct.\,20/21&240&10&2400& \\
          &blue&Oct.\,21/22&240& 5&1200&seeing $\sim$1\arcsecpoint4, clouds\\
          &red &Oct.\,21/22&240&12&2880&seeing $\sim$1\arcsecpoint6\\
Q 2348-4025&blue&Oct.\,20/21&240& 6&1440&seeing $\sim$2\arcsecpoint0\\
          &red &Oct.\,20/21&240& 6&1440&seeing $\sim$1\arcsecpoint5\\
          &blue&Oct.\,21/22&240& 6&1440&seeing $\sim$1\arcsecpoint7,clouds\\
          &red &Oct.\,21/22&240&12&2880&seeing $\sim$1\arcsecpoint7,clouds\\
          &blue&Oct.\,22/23&180&16&2880&clouds\\
\hline
HD 19904   &blue&Oct.\,19/20&  5& 4&  20&seeing $\sim$0\arcsecpoint5\\
          &red &Oct.\,19/20&  5& 4&  20& \\
          &blue&Oct.\,20/21&  5& 4&  20&seeing $\sim$1\arcsecpoint5\\ 
          &red &Oct.\,20/21&  5& 4&  20& \\
HD 25402   &blue&Oct.\,19/20&  5& 4&  20& \\
          &red &Oct.\,19/20&  5& 4&  20&seeing $\sim$0\arcsecpoint6\\
          &blue&Oct.\,20/21&  5& 4&  20&seeing $\sim$1\arcsecpoint2\\
          &red &Oct.\,20/21&  5& 4&  20& \\
HD 38921   &blue&Oct.\,19/20&  5& 4&  20&seeing $\sim$0\arcsecpoint6\\
          &red &Oct.\,19/20&  5& 4&  20& \\
HD 205772  &blue&Oct.\,19/20& 10& 8&  80& \\
          &blue&Oct.\,19/20&  5& 4&  20& \\
          &red &Oct.\,19/20&  5& 4&  20& \\ 
          &blue&Oct.\,20/21&  5& 4&  20& \\
          &red &Oct.\,20/21&  5& 4&  20& \\
          &blue&Oct.\,21/22&  5& 4&  20&clouds\\
          &red &Oct.\,21/22&  5& 4&  20&clouds\\
          &blue&Oct.\,22/23&  5& 8&  40& \\
HD 210395  &blue&Oct.\,19/20&  5& 4&  20& \\
          &red &Oct.\,19/20&  5& 4&  20& \\
          &blue&Oct.\,20/21&  5& 4&  20& \\
          &red &Oct.\,20/21&  5& 4&  20& \\
          &blue&Oct.\,21/22&  5& 8&  40&clouds\\
          &red &Oct.\,21/22&  5& 8&  40&clouds\\
\enddata
\end{deluxetable}


\begin{deluxetable}{lcccc}
\tablewidth{0pt}
\tablecaption{total and effective integration times}
\tablehead{
\colhead{quasar} &
\colhead{t$_{tot}^{blue}$} &
\colhead{t$_{eff}^{blue}$} &
\colhead{t$_{tot}^{red}$} &
\colhead{t$_{eff}^{red}$} \\
\colhead{(1)} &
\colhead{(2)} &
\colhead{(3)} &
\colhead{(4)} &
\colhead{(5)} 
}
\startdata
Q 0103-260 & 4800&4800&6720&4560\\
Q 0105-2634& 2880&2880&1440&1440\\
Q 0256-0000& 5400&2880&3420&2880\\
Q 0302-0019& 5760&5760&6480&6000\\
Q 2227-3928& 5520&4320&6720&5760\\
Q 2348-4025& 5760&3720&4320&2160\\
\enddata
\end{deluxetable}

\clearpage



\ptlandscape
\begin{deluxetable}{lccccccc}
\rotate
\tablewidth{215mm}
\tablecaption{emission line measurements}
\tablehead{
\colhead{feature} &
\colhead{Q 0103-260}& 
\colhead{Q 0105-2634}&
\colhead{Q 0256-0000}&
\colhead{Q 0302-0019}&
\colhead{Q 2227-3928}&
\colhead{Q 2348-4025}&
\colhead{local QSO}
}
\startdata
\cii ]2326                 &    ---      &   ---       &    ---      &    ---      &   ---       &    ---      &0.05$\pm$0.01\\
{[}\neiv ]2423             &    ---      &   ---       &0.05$\pm$0.01&    ---      &   ---       &    ---      &    ---      \\
\mgii 2798                 &1.00$\pm$0.05&1.00$\pm$0.04&1.00$\pm$0.03&1.00$\pm$0.03&1.00$\pm$0.03&1.00$\pm$0.06&1.00$\pm$0.05\\
{[}\neiii ]3869            &    ---      &    ---      &    ---      &0.13$\pm$0.02&    ---      &    ---      &    ---      \\
H$\delta$4101              &    ---      &    ---      &0.34$\pm$0.04&0.26$\pm$0.04&    ---      &    ---      &    ---      \\
H$\gamma $4340             &0.52$\pm$0.11&0.37$\pm$0.06&    ---      &0.34$\pm$0.05&0.43$\pm$0.08&    ---      &    ---      \\
H$\beta $4861              &1.84$\pm$0.13&0.84$\pm$0.07&1.32$\pm$0.08&1.01$\pm$0.05&1.12$\pm$0.07&0.73$\pm$0.08&    ---      \\
{[}\oiii ]4959             &0.03:        &    ---      &0.06$\pm$0.01&0.18$\pm$0.01&    ---      &    ---      &    ---      \\   
{[}\oiii ]5007             &0.10:        &    ---      &0.22$\pm$0.02&0.59$\pm$0.02&0.015:       &    ---      &    ---      \\
\hline													       
\feii 4570 \tablenotemark{a}&0.99$\pm$0.16&1.01$\pm$0.13&0.57$\pm$0.06&0.58$\pm$0.07&0.64$\pm$0.09&0.78$\pm$0.10&    ---     \\
\feii 4924,5018\tablenotemark{b}&0.36$\pm$0.06&0.37$\pm$0.05&0.21$\pm$0.02&0.22$\pm$0.03&0.24$\pm$0.03&0.28$\pm$0.04&  ---   \\
\feii 5190 \tablenotemark{c}&0.58$\pm$0.09&0.59$\pm$0.08&0.33$\pm$0.03&0.34$\pm$0.04&0.38$\pm$0.05&0.46$\pm$0.06&    ---     \\
\feii \,opt\tablenotemark{d}&1.95$\pm$0.31&1.99$\pm$0.26&1.12$\pm$0.11&1.15$\pm$0.14&1.26$\pm$0.17&1.53$\pm$0.19& ---     \\
\hline													       
\feii 2080 \tablenotemark{e}&    ---      &    ---      &0.18$\pm$0.02&    ---      &    ---      &    ---      &0.23$\pm$0.02\\
\feii 2500 \tablenotemark{f}&2.43$\pm$0.29&2.26$\pm$0.26&1.73$\pm$0.15&2.44$\pm$0.23&2.29$\pm$0.24&1.94$\pm$0.23&2.25$\pm$0.23\\
\feii 2680 \tablenotemark{g}&0.59$\pm$0.07&0.54$\pm$0.06&0.42$\pm$0.04&0.59$\pm$0.06&0.55$\pm$0.06&0.47$\pm$0.06&0.54$\pm$0.06\\
\feii 2900 \tablenotemark{h}&1.03$\pm$0.12&0.95$\pm$0.11&0.73$\pm$0.06&1.03$\pm$0.10&0.96$\pm$0.10&0.82$\pm$0.10&0.95$\pm$0.10\\
\feii UV \tablenotemark{i}&4.14$\pm$0.50&3.84$\pm$0.44&2.95$\pm$0.26&4.16$\pm$0.40&3.90$\pm$0.40&3.31$\pm$0.38&3.82$\pm$0.40\\
\hline													       
\feii \,UV/\feii \,opt &2.13$\pm$0.40&1.93$\pm$0.33&2.64$\pm$0.33&3.62$\pm$0.53&3.10$\pm$0.51&2.16$\pm$0.32&    ---      \\
\feii \,UV $+$ \feii \,opt&6.09$\pm$0.68&5.83$\pm$0.63&4.07$\pm$0.36&5.31$\pm$0.50&5.15$\pm$0.53&4.84$\pm$0.55& --- \\
Balmer cont. \tablenotemark{j}&3.86$\pm$0.60&5.45$\pm$0.60&5.15$\pm$0.50&5.40$\pm$0.49&1.37$\pm$0.14&6.16$\pm$0.67&5.02$\pm$0.51\\
$\tau _{BaC}$              &1.0&0.5&0.5&1.0&1.0&1.0&1.0\\
\hline
$\alpha$&$-$1.33$\pm$0.06&$-$0.69$\pm$0.06&$-$0.41$\pm$0.05&$-$0.21$\pm$0.03&$-$0.85$\pm$0.05&$-$0.53$\pm$0.07&$-$0.34$\pm$0.03\\
I$_{rest}$(\mgii $\lambda 2798$)&73.6$\pm$3.5&267.4$\pm$11.0&200.7$\pm$7.0&341.5$\pm$10.0&156.0$\pm$5.0&155.6$\pm$10.0&317.2$\pm$16.7\\
{[}$10^{-15}$erg\,s$^{-1}$\,cm$^{-2}$]& & & & & & & \\
\enddata
\tablenotetext{a}{$\lambda \lambda 4250 - 4770$ \AA ,
$^b \lambda \lambda 4800 - 5085$ \AA ,
$^c \lambda \lambda 5085 - 5500$ \AA ,
$^d \lambda \lambda 4250 - 5500$ \AA ,
$^e \lambda \lambda 2030 - 2130$ \AA ,
$^f \lambda \lambda 2240 - 2660$ \AA ,
$^g \lambda \lambda 2660 - 2790$ \AA ,
$^h \lambda \lambda 2790 - 3030$ \AA ,
$^i \lambda \lambda 2200 - 3090$ \AA ,
$^j \lambda \lambda 2200 - 3650$ \AA }
\end{deluxetable}

\end{document}